\documentclass{llncs}
\usepackage{epsf}

\begin{document}

\title{Weight vs Magnetization Enumerator for Gallager Codes}
\author{Jort van~Mourik$^1$, David Saad$^1$ and Yoshiyuki Kabashima$^2$}
\institute{
$^1$~Neural Computing Research Group,  Aston University,
Birmingham B4 7ET, UK \\
$^2$~Department of Computational Intelligence and Systems Science,
Tokyo Institute of Technology, Yokohama 2268502, Japan
}

\newcommand {\e   } {\!+\!                  }
\newcommand {\m   } {\!-\!                  }
\newcommand {\Bra } {\left\langle           }
\newcommand {\Ket } {\right\rangle          }
\newcommand {\beq } {\begin{equation}} \newcommand {\eeq } {\end{equation}}
\newcommand {\bea } {\begin{eqnarray}} \newcommand {\eea } {\end{eqnarray}}
\newcommand {\hsc } {\hspace*{1cm}   }
\newcommand {\pa  } {\partial               }
\newcommand {\lh  } {\left(                          }
\newcommand {\rh  } {\right)                         }
\newcommand {\lv  } {\left[                          }
\newcommand {\rv  } {\right]                         }
\newcommand {\be  } {\beta      }
\newcommand {\de  } {\delta     }
\newcommand {\om  } {\omega     }
\newcommand {\la  } {\lambda    }
\newcommand {\htom} {{\hat{\om }}}
\newcommand {\htpi} {{\hat{\pi }}}
\newcommand {\htm } {{\hat{m   }}}
\newcommand {\htq } {{\hat{q   }}}
\newcommand {\htx } {{\hat{x   }}}
\newcommand {\cF  } {{\cal   F }}       
\newcommand {\cI  } {{\cal   I }}
\newcommand {\cM  } {{\cal   M }}       
\newcommand {\cQ  } {{\cal   Q }}       
\newcommand {\cS  } {{\cal   S }}      
\newcommand {\cU  } {{\cal   U }}      
\newcommand {\cW  } {{\cal   W }}      
\newcommand {\cZ  } {{\cal   Z }}      
\newcommand {\bfA } {{\bf    A }}
\newcommand {\bfC } {{\bf    C }}
\newcommand {\bfG } {{\bf    G }}
\newcommand {\vs  } {{\vec{  s }}}
\newcommand {\vn  } {{\vec{  n }}}
\newcommand {\vr  } {{\vec{  r }}}
\newcommand {\vt  } {{\vec{  t }}}
\newcommand {\vz  } {{\vec{  z }}}
\newcommand {\nn  } {\nonumber  }
\newcommand {\Tr  } {\mathop{\mbox{\rm Tr}}   }
\newcommand {\fns } {\footnotesize}
\newcommand {\ov  } {\over         }
\newcommand {\ha  } {{1\over 2}}
\newsavebox{\uuunit}
\sbox{\uuunit}
    {\setlength{\unitlength}{0.825em}
     \begin{picture}(0.6,0.7)
        \thinlines
        \put(0,0){\line(1,0){0.5}}
        \put(0.15,0){\line(0,1){0.7}}
        \put(0.35,0){\line(0,1){0.8}}
       \multiput(0.3,0.8)(-0.04,-0.02){12}{\rule{0.5pt}{0.5pt}}
     \end {picture}}
\newcommand {\Unity}{\mathord{\!\usebox{\uuunit}}}
\maketitle
\begin{abstract}
  We propose a method to determine the critical noise level for
  decoding Gallager type low density parity check error correcting
  codes. The method is based on the magnetization enumerator ($\cM$),
  rather than on the weight enumerator ($\cW$) presented recently in
  the information theory literature.  The interpretation of our method
  is appealingly simple, and the relation between the different
  decoding schemes such as typical pairs decoding, MAP, and finite
  temperature decoding (MPM) becomes clear.  Our results are more
  optimistic than those derived via the methods of information theory
  and are in excellent agreement with recent results from another
  statistical physics approach.
\end{abstract}

\section{Introduction}
%
Triggered by active investigations on error correcting codes in both of
information theory (IT)~\cite{MacKay,Richardson,Aji} and statistical physics 
(SP)~\cite{us_PRL,NishimoriWong} communities, there is a growing interest
in the relationship between IT and SP. As  the two communities investigate  
similar problems, one may expect that standard techniques known in one framework
would bring about new developments in the other, and vice versa.
Here we present a direct SP method to determine the critical noise level for
Gallager type low density parity check codes which allows us to focus on
the differences between the various decoding criteria and their approach for
defining the critical noise level for which decoding, using Low Density Parity
Check (LDPC) codes, is theoretically feasible.
%
\section{Gallager code}

In a general scenario, the $N$ dimensional Boolean message $\vs^{o}\in\{0,1\}^N$
is encoded to the $M(\!>\!N)$ dimensional Boolean vector $\vt^o$, and
transmitted via a noisy channel, which is taken here to be a Binary Symmetric
Channel (BSC) characterized by an independent flip probability $p$ per bit;
other transmission channels may also be examined within a similar framework. 
At the other end of the channel, the corrupted codeword is decoded
utilizing the structured codeword redundancy.

The error correcting code that we focus on here is Gallager's linear
code~\cite{Gallager}. Gallager's code is a low density parity check code defined
by the a binary $(M\m N)\!\times\!M$ matrix $\bfA=[\bfC_1|\bfC_2]$,
concatenating two very sparse matrices known to both sender and receiver, with
the $(M\m N)\!\times\!(M\m N)$ matrix $\bfC_2$ being invertible. The matrix
$\bfA$ has $K$ non-zero elements per row and $C$ per column, and the code rate
is given by $R\!=\!1\m C/K\!=\!1\m N/M$. Encoding refers to multiplying the
original message $\vs ^o$ with the $(M\!\times \!N)$ matrix ${\bf G}^T$ (where
${\bf G}\!=\! [\mbox{\boldmath{$\Unity$}}_N|{\bf C}_2^{\m 1}]$), yielding the
transmitted vector  $\vt ^o$. Note that all operations are carried out in (mod
2) arithmetic. Upon sending $\vec{t}^o$ through the binary symmetric channel
(BSC) with noise level $p$, the vector $\vr=\vt ^o\e \vn ^o$ is received, where
$\vn ^o$ is the true noise.

Decoding is carried out by multiplying $\vr $ by $\bfA$ to produce the syndrome
vector $\vz\!=\!\bfA \vr $ ($=\bfA \vn ^o$, since $\bfA \bfG ^{T}={\bf 0}$).
In order to reconstruct the original message $\vs ^o$, one has to obtain an
estimate $\vn $ for the true noise $\vn ^o$. First we select all $\vn $ that
satisfy the parity checks $\bfA \vn =\bfA \vn ^o$:
\begin{equation}
\cI_{\rm pc}(\bfA ,\vn ^o)\equiv\{\vn ~|~\bfA \vn =\vz \}, \  \mbox{and} \
\cI^{\rm r}_{\rm pc}(\bfA ,\vn ^o)\equiv\{\vn \in \cI_{\rm pc}(\bfA,\vn ^o)|\vn
\neq \vn ^o\},
\label{Ipc}
\end{equation}
the (restricted) parity check set. Any general decoding scheme then consists of
selecting a vector $\vn ^*$ from $\cI_{\rm pc}({\bf A},\vn ^o)$ on the basis of
some noise statistics criterion. Upon successful decoding $\vn ^0$ will be
selected, while a decoding error is declared when a vector $\vn ^*\in \cI^{\rm
r}_{\rm pc}(\bfA ,\vn ^o)$ is selected. An measure for the error probability is
usually defined in the information theory literature~\cite{gallager_book} as
\beq 
P_e(p)=\Bra \Delta\left(~\exists~\vn \in \cI^{\rm r}_{\rm pc}(\bfA ,\vn ^o)
:w(\vn )\le w(\vn^o)~|~\vn ^o\right)~\Ket _{\bfA ,\vn ^o} \ ,
\label{Pe}
\eeq 
where  $\Delta(\cdot)$ is an indicator function returning 1 if there exists a 
vector $\vn\in \cI^{\rm r}_{\rm pc}(\bfA ,\vn ^o)$ with lower weight than that
of the given noise vector $\vn ^o$. The weight of a vector is the average sum of
its components $w(\vn )\equiv{1\over M}\sum_{j=1}^Mn_j$. To obtain the error
probability, one averages the indicator function over all  $\vn ^o$ vectors
drawn from some distribution and the code ensemble $\bfA$ as denoted by
$\Bra.\Ket _{\bfA,\vn ^o}$.

Carrying out averages over the indicator function is difficult, and the error 
probability (\ref{Pe}) is therefore upper-bounded by averaging over the {\em
number} of vectors $\vn$ obeying the weight condition $w(\vn )\ge w(\vn^o)$.
Alternatively, one can find the average number of vectors with a given weight 
value $w$ from which one can construct a complete weight distribution of noise
vectors  $\vn$ in  $\cI^{\rm r}_{\rm pc}(\bfA ,\vn ^o)$. From this distribution
one can, in principle, calculate a bound for $P_e$ and derive critical noise
values above which successful decoding cannot be carried out.

A natural and direct measure for the average number of states is the entropy 
of a system under the restrictions described above, that can be calculated 
via the methods of statistical physics.

It was previously shown (see e.g. \cite{us_PRL} for technical details) that this
problem can be cast into a statistical mechanics formulation, by replacing the
field $(\{0,1\},+{\rm mod(2)})$ by ($\{1,-1\},\times$), and by adapting the
parity checks correspondingly. The statistics of a noise vector $\vn $ is now
described by its magnetization $m(\vn )\equiv{1\over M}\sum_{j=1}^Mn_j$,
$(m(\vn)\in [1,-1])$, which is inversely linked to the vector weight in the
$[0,1]$ representation. With this in mind, we introduce the conditioned
magnetization enumerator, for a given  code and noise, measuring the noise
vector magnetization distribution in $\cI^{\rm r}_{\rm pc}({\bf A},\vn^o)$
\beq
\cM_{\bfA ,\vn ^o}(m)\equiv{1\over M}\ln\left[
\Tr_{\vn \in\cI ^{\rm r}_{\rm pc}(\bfA ,\vn ^o)}\delta(m(\vn )\m m)\right]~.
\label{MAn}
\eeq
To obtain the {\em magnetization enumerator} $\cM(m)$
\beq
\cM(m)=\Bra \frac{}{}~\cM_{\bfA ,\vn ^o}(m)~\Ket _{\bfA ,\vn ^o}~,
\label{M}
\eeq 
which is the entropy of the noise vectors in $\cI^{\rm r}_{\rm pc}(\bfA,\vn^0)$
with a given $m$, one carries out  uniform explicit averages over all codes
$\bfA$ with given parameters $K,C$, and  weighted average over all possible
noise vectors generated by the BSC, i.e., 
\beq
P(\vn^o)=\prod_{j=1}^M\left((1\m p)~\delta(n^o_j\m1)+p~\delta(n^o_j\e1)\right).
\eeq
It is important to note that, in calculating the entropy, the average quantity 
of interest is the magnetization enumerator rather than the actual number of
states. For physicists, this is the natural way to carry out the averages due to
three main reasons: a) The entropy obtained in this way is believed to be {\em 
self-averaging}, i.e., its average value (over the disorder) coincides with its
{\em typical} value. b) This quantity is {\em extensive} and grows linearly with
the system size. c) This averaging distinguishes between {\em annealed}
variables that are averaged or summed for a given set of {\em quenched}
variables, that are averaged over later on. In this particular case, summation
over all $\vn$ vectors is carried for a {\em fixed} choice of code $\bfA$ and
noise vector $\vn^o$; averages over these variables are carried out at the
next level.

One should point out that in somewhat similar calculations, we showed that this
method of carrying out the averages provides more accurate results in comparison
to averaging over both sets of variables simultaneously~\cite{reliability}.

A positive magnetization enumerator, $\cM(m)\!>\!0$ indicates that there is an
exponential number of solutions (in $M$) with magnetization $m$, for typically
chosen $\bfA$ and $\vn^o$,  while $\cM(m)\!\to\!0$ indicates that this number
vanishes as $M\!\to\!\infty$ (note that negative entropy is unphysical in
discrete systems).

Another important indicator for successful decoding is the overlap $\om$ between
the selected estimate $\vn ^*$, and the true noise $\vn ^o$:~$\om(\vn,\vn^o)
\equiv{1\ov M}\sum_{j=1}^Mn_jn^o_j$, $(\om(\vn,\vn^o)\in [-1,1])$, with $\om=1$
for successful (perfect) decoding. However, this quantity cannot be used for
decoding as $\vn ^o$ is unknown to the receiver. The (code and noise dependent)
overlap enumerator is now defined as:
\beq
\cW _{\bfA ,\vn ^o}(\om )\equiv{1\ov M}\ln\left[\Tr_{\vn \in\cI^{\rm r}_{\rm
pc}(\bfA ,\vn ^o)}\de (\om(\vn ,\vn ^o)\m \om )\right] \ ,
\eeq
and the average quantity being 
\beq 
\cW(\om)=\Bra\frac{}{}\cW_{\bfA,\vn^o}(\om)\Ket_{\bfA ,\vn ^o} \ .
\eeq
This measure is directly linked to the {\em weight enumerator} \cite{Aji}),
although according to our notation, averages are carried out distinguishing
between annealed and quenched variables unlike the common definition in the IT
literature. However, as we will show below, the two types of averages provide
identical results {\em in this particular case}.
%
\section{The statistical physics approach}

Quantities of the type $\cQ(c)=\Bra\cQ_y(c)\Ket_y$, with
$\cQ_y(c)={1\ov M}\ln\left[{\cal Z}_y(c)\right]$ and
$\cZ_y(c)\equiv\Tr_x~\delta(c(x,y)\m Mc)$,  are very common in the SP of
disordered systems; the macroscopic order parameter $c(x,y)$ is fixed to a
specific value and may depend both on the disorder $y$ and on the microscopic
variables $x$. Although we will not prove this here, such a quantity is
generally believed to be {\em self-averaging} in the large system limit, i.e.,
obeying a probability distribution  $P\left(\cQ_y(c))=\delta(\cQ_y(c)-\cQ(c))
\right)$. The direct calculation of $\cQ(c)$ is known as a {\em quenched}
average over the disorder, but is typically hard to carry out and requires using
the replica method~\cite{nishimori_book}. The replica method makes use of the
identity $\Bra\ln\cZ\Ket=\Bra\ \lim_{n\to 0}[\cZ^n\m 1]/n\ \Ket$,
by calculating averages over a product of partition function replicas. Employing
assumptions about replica symmetries and analytically continuing the variable
$n$ to zero, one obtains solutions which enable one to determine the state of
the system.

To simplify the calculation, one often employs the so-called {\em annealed}
approximation, which consists of performing an average over $\cQ_y(c)$ first,
followed by the logarithm operation. This avoids the replica method and provides
(through the convexity of the logarithm function) an upper bound to the quenched
quantity:
\beq
Q_a(c)\equiv{1\ov M}\ln[\Bra \cZ_y(c)\Ket _y]~\geq~
Q_q(c)\equiv{1\ov M}\Bra \ln[\cZ_y(c)]\Ket _y=\lim_{n\!\to\!0}
{\Bra\cZ_y^n(c)\Ket_y\m 1\ov nM}~.
\label{Qaq}
\eeq

The full technical details of the calculation will be presented elsewhere, and
those of a very similar calculation can be found in e.g. \cite{us_PRL}. It turns
out that it is useful to perform the gauge transformation $n_j\!\to\!n_jn^o_j$,
such that the averages over the code $\bfA $ and noise $\vn ^o$ can be
separated, $\cW_{\bfA ,\vn ^o}$ becomes independent of $\vn ^o$, leading to an
equality between the quenched and annealed results, $\cW(m)=\cM_a(m)|_{p=0}=
\cM_q(m)|_{p=0}$. For any finite noise value $p$ one should multiply 
$\exp[\cW(\om)]$ by the probability that a state obeys all parity checks
$\exp[-K(\om,p)]$ given an overlap $\om$ and a noise level $p$~\cite{Aji}. In
calculating $\cW(\om)$ and $\cM_{a/q}(m)$, the $\de$-functions fixing $m$ and
$\om$, are enforced by introducing Lagrange multipliers $\htm$ and $\htom$.

Carrying out the averages explicitly one then employs the saddle point method to
extremize the averaged quantity with respect to the parameters introduced while 
carrying out the calculation. These lead, in both quenched and annealed
calculations, to a set of saddle point equations that are solved either
analytically or numerically to obtain the final expression for the averaged
quantity (entropy).

The final expressions for the annealed entropy, under both overlap ($\om$) and
magnetization ($m$) constraints, are of the form:
\bea
\cQ_a&=&-{C\ov K}\!\left(\ln(2)\e(K\!\m\!1)\ln(1\e q_1^K)\right)\e\ln\!\Bra
\Tr_{n=\pm1}{\rm e}^{(n\htom \e\htm n^o)}(1\e nq_1^{K\m1})^C\Ket_{n^o}
\nonumber \\
&&- \htom\om - \htm m~,
\label{Qa}
\eea 
where $q_1$ has to be obtained from the saddle point equation ${\partial\cQ_a
\ov\partial q_1}=0$. Similarly, the final expression in the quenched
calculation, employing the simplest replica symmetry
assumption~\cite{nishimori_book}, is of the form:
\bea
\cQ _q\!&=&-\!C\!\!\int\!\!dxd\htx \ \pi(x)\htpi(\htx)~\ln[1\e x\htx]\e
{C\ov K}\int\!\left\{\prod_{k=1}^Kdx_k\pi(x_k)\right\}\ln\left[\ha
\left(1\e\!\prod_{k=1}^K\!x_k\right)\right]\nonumber\\
&&+\!\int\!\left\{\prod_{c=1}^Cd\htx_c\htpi 
(\htx_c)\right\}\Bra\!\ln\left[\Tr_{n=\pm1}\exp(n(\hat{\om}\e\hat{m}n^o))
\prod_{c=1}^C(1\e n\htx_c) 
\right]\!\Ket_{n^o} \nonumber\\
&& - \htom\om - \htm m~. 
\label{Qq}
\eea
The probability distributions $\pi(x)$ and $\htpi(\htx)$ emerge from the
calculation; the former represents a probability distribution with respect to
the noise vector local magnetization~\cite{TAP_book}, while the latter relates
to a field of conjugate variables which emerge from the introduction of
$\de$-functions while carrying out the averages (for details see \cite{us_PRL}).
Their explicit forms are obtained from the functional saddle point equations
${\de\cQ_q\over\de \pi(x)}$, ${\de{\cQ}_q\ov\de\htpi(\htx)}=0$, and all
integrals are from $\m1$ to 1. Enforcing a $\de $-function corresponds to taking
$\htom ,\htm $ such that ${\partial\cQ_{a/q}\ov\partial\htom},{\partial\cQ_{a/q}
\ov\partial\htm}=0$, while not enforcing it corresponds to putting $\htom ,\htm
$ to 0. Since $\om$, $m$ follow from ${\partial\cQ_{a/q}\ov\partial\htom}\!=
\!0$, ${\partial\cQ_{a/q}\ov\partial\htm}\!=\!0$, all the relevant quantities
can be recovered with appropriate choices of $\htom$, $\htm$.
\begin{figure}[h]
\setlength{\unitlength}{0.88mm}
\begin{picture}(140,100)
\put(  0, 50){\epsfysize=50\unitlength\epsfbox{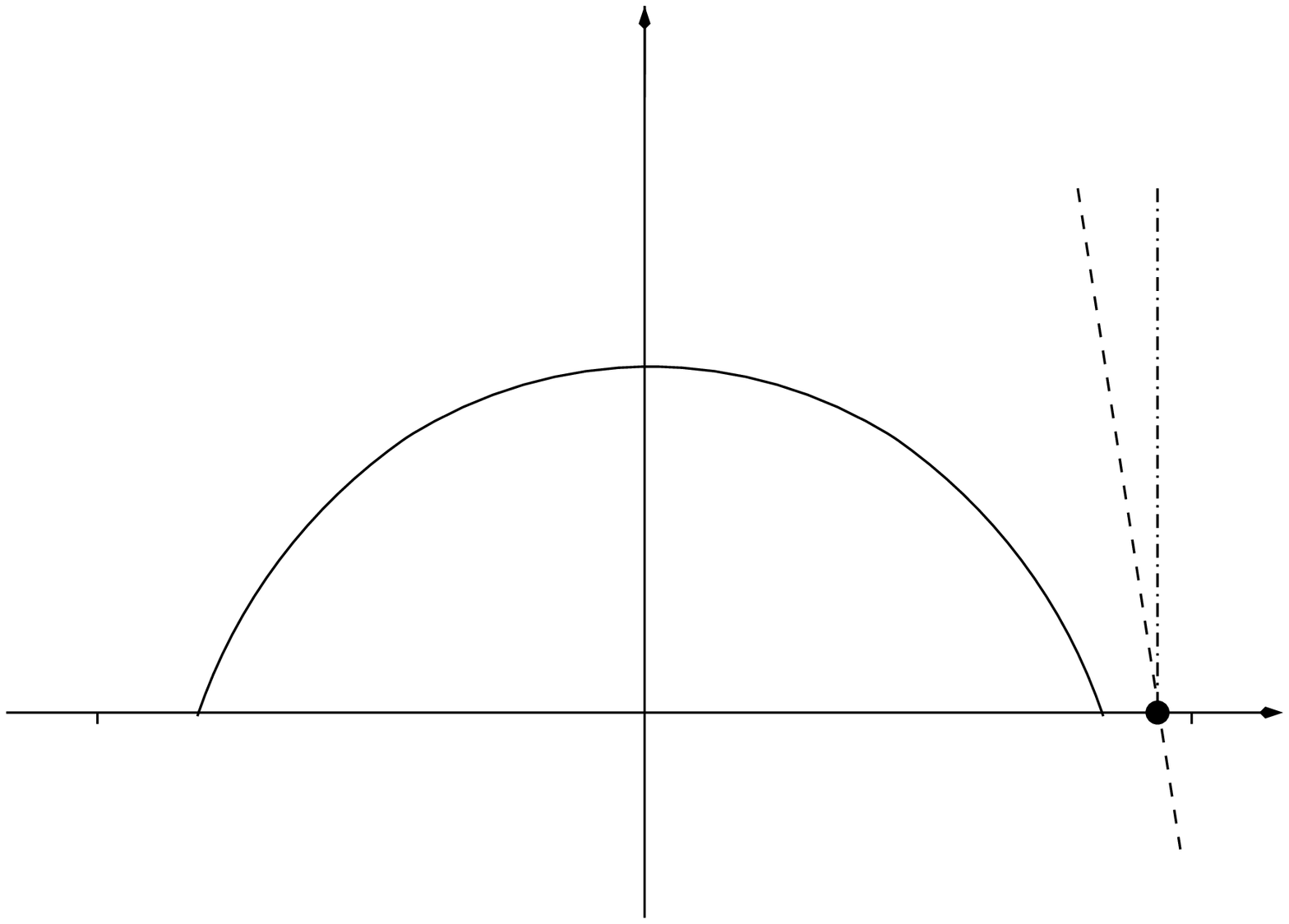}}
\put(  5, 90){\mbox{\fns \boldmath ${\rm a)}~p\!<\!p_c$}}
\put( 35, 95){\mbox{\fns \boldmath $\cM(m)$}}
\put( 40, 58){\mbox{\fns \boldmath $m$}}
\put( 50, 58){\mbox{\fns \boldmath $m_{\e}(p)$}}
\put(  2, 58){\mbox{\fns \boldmath $-1$}}
\put( 64, 58){\mbox{\fns \boldmath $1$}}
\put( 70, 50){\epsfysize=50\unitlength\epsfbox{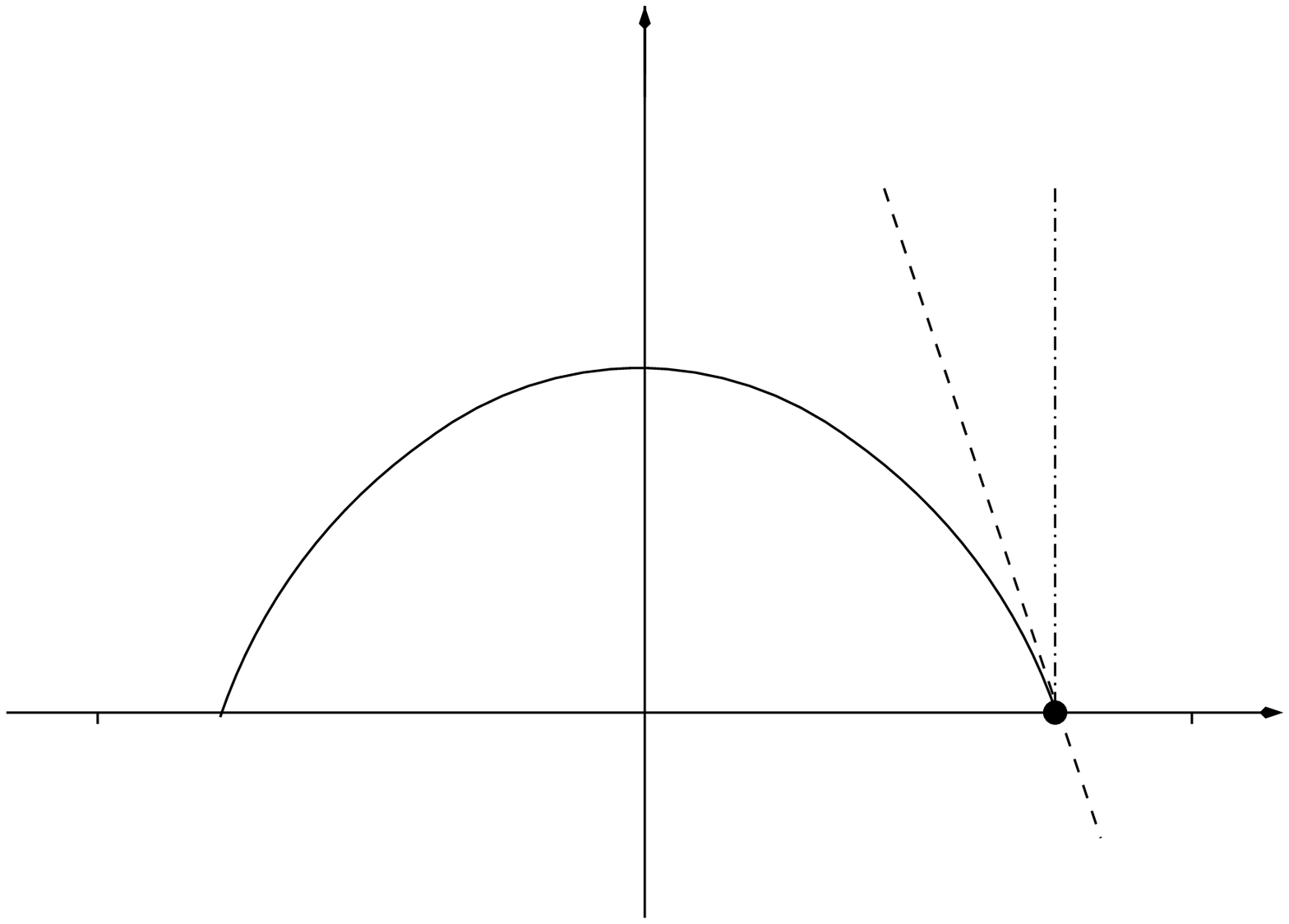}}
\put( 75, 90){\mbox{\fns \boldmath ${\rm b)}~p\!=\!p_c$}}
\put(105, 95){\mbox{\fns \boldmath $\cM(m)$}}
\put(110, 58){\mbox{\fns \boldmath $m$}}
\put(120, 58){\mbox{\fns \boldmath $m_{\e}(p)$}}
\put( 72, 58){\mbox{\fns \boldmath $-1$}}
\put(134, 58){\mbox{\fns \boldmath $1$}}
\put( 35,  0){\epsfysize=50\unitlength\epsfbox{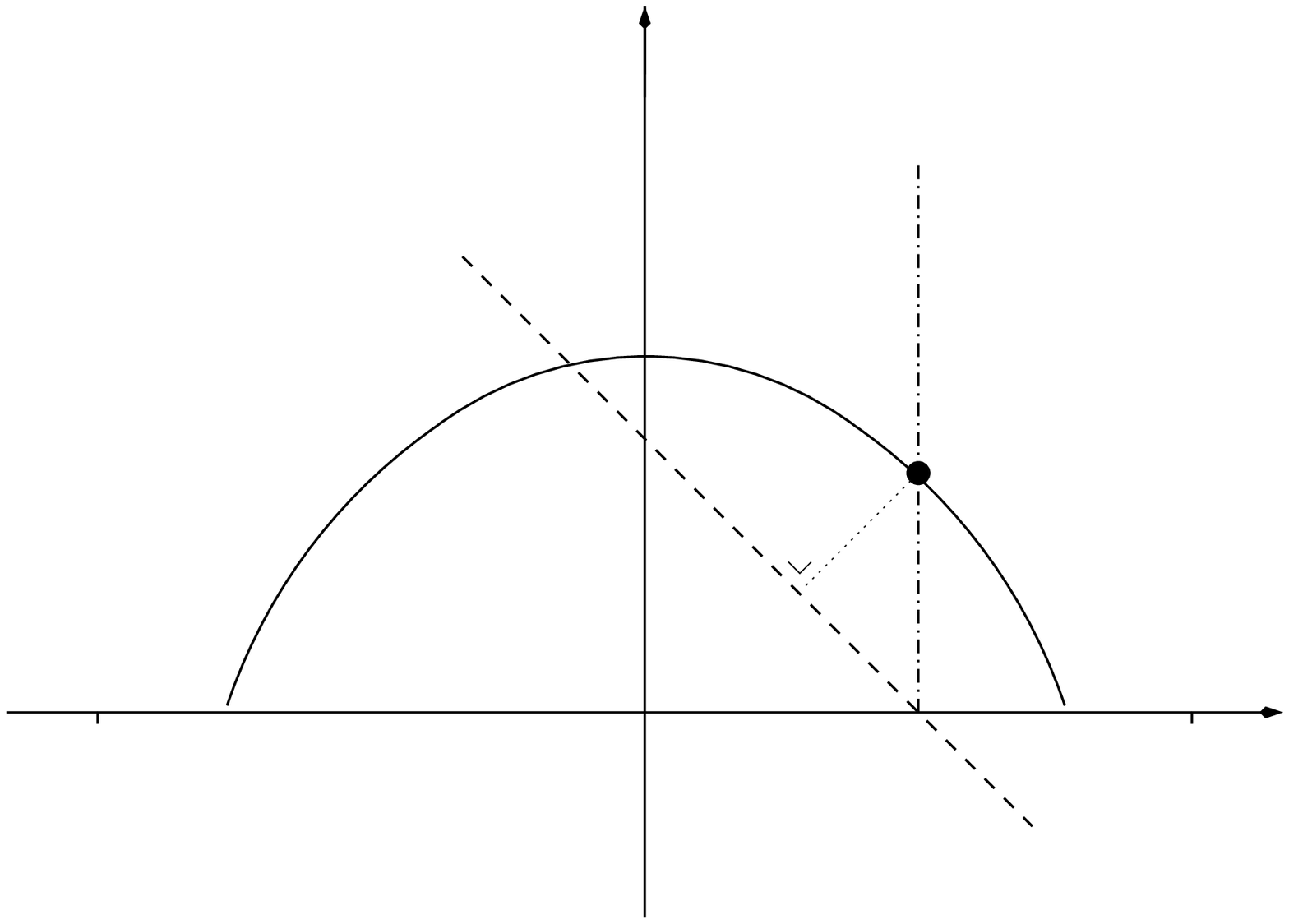}}
\put( 40, 40){\mbox{\fns \boldmath ${\rm c)}~p\!>\!p_c$}}
\put( 70, 45){\mbox{\fns \boldmath $\cM(m)$}}
\put( 75,  8){\mbox{\fns \boldmath $m$}}
\put( 85,  8){\mbox{\fns \boldmath $m_{\e}(p)$}}
\put( 37,  8){\mbox{\fns \boldmath $-1$}}
\put( 99,  8){\mbox{\fns \boldmath $1$}}
\put( 63,80){\mbox{\fns \boldmath $m_0(p)$}}
\put(127,80){\mbox{\fns \boldmath $m_0(p)$}}
\put( 85,30){\mbox{\fns \boldmath $m_0(p)$}}
\end{picture}
\caption{
  The qualitative picture of $\cM(m)\!\geq\!0$ (solid lines) for
  different values of $p$. For MAP, MPM and typical set decoding, only
  the relative values of $m_{\e}(p)$ and $m_{0}(p)$ determine the
  critical noise level. Dashed lines correspond to the energy contribution of
  $-\be F$ at Nishimori's condition ($\be=1$). The states with the lowest free
  energy are indicated with $\bullet$.
  {\bf a)} Sub-critical noise levels $p\!<\!p_{c}$, where $m_{\e}(p)\!<\!m_{0}
  (p)$, there are no solutions with higher magnetization than $m_{0}(p)$, and
  the correct solution has the lowest free energy.
  {\bf b)} Critical noise level $p\!=\!p_c$, where $m_{\e}(p)\!=\!m_{0}(p)$. The
  minimum of the free energy of the sub-optimal solutions is equal to that of
  the correct solution at Nishimori's condition.
  {\bf c)} Over-critical noise levels $p\!>\!p_c$ where many solutions have a
  higher magnetization than the true typical one. The minimum of the free
  energy of the sub-optimal solutions is lower than that of the
  correct solution.
}\label{fig:1}
\end{figure}
%
\section{Qualitative picture}
%
We now discuss the qualitative behaviour of $\cM(m)$, and the interpretation of
the various decoding schemes. To obtain separate results for  $\cM(m)$ and
$\cW(m)$ we calculate the results of Eqs.(\ref{Qa}) and (\ref{Qq}),
corresponding to the annealed and quenched cases respectively, setting $\htom=0$
for obtaining $\cM(m)$ and $\htm\!=\!0$ for obtaining $\cW(\om)$ (that becomes
$\cW(m)$ after gauging). In Fig. \ref{fig:1}, we have qualitatively plotted the
resulting function $\cM(m)$ for relevant values of $p$.
$\cM(m)$ (solid line) only takes positive values in the interval $[m_{\m}(p),
m_{\e}(p)]$; for even $K$, $\cM(m)$ is an even function of $m$ and
$m_{\m}(p)\!=\!-m_{\e}(p)$. The maximum value of $\cM(m)$ is always $(1\m
R)\ln(2)$. The true noise $\vn^o$ has (with probability 1) the typical
magnetization of the BSC: $m(\vn^o)\!=\!m_0(p)\!=\!1\m2p$ (dashed-dotted line).

The various decoding schemes can be summarized as follows:
\begin{itemize}
\item{\bf Maximum likelihood (MAP) decoding} -
  minimizes the {\em block error probability}~\cite{iba} and consists of
  selecting the $\vn $ from $\cI_{\rm pc}(\bfA ,\vn ^0)$ with the highest
  magnetization. Since the probability of error below $m_{\e}(p)$ vanishes,
  $P(\exists\vn\in\cI^{\rm r}_{\rm pc}: m(\vn )\!>\! m_{\e}(p))\!=\!0$, and
  since $P(m(\vn^o)\!=\!m_0(p))\!=\!1$, the critical noise level $p_c$ is
  determined by the condition $m_{\e}(p_c)\!=\!m_0(p_c)$. The selection process
  is explained in Fig.\ref{fig:1}(a)-(c).
\item{\bf Typical pairs decoding} -
  is based on randomly selecting a $\vn $ from $\cI_{\rm pc}$ with $m(\vn)=m_0
  (p)$~\cite{Aji}; an error is declared when $\vn ^0$ is not the only element of
  $\cI_{\rm pc}$. For the same reason as above, the critical noise level $p_c$
  is determined by the condition $m_{\e}(p_c)\!=\!m_0(p_c)$.  
\item{\bf Finite temperature (MPM) decoding} - An energy $-Fm(\vn )$
  (with $F\!=\!\ha\ln({1-p\ov p})$) according to Nishimori's condition\footnote{
  This condition corresponds to the selection of an accurate prior
  within the Bayesian framework.}
  is attributed to each $\vn \in\cI_{\rm pc}$, and a solution is chosen from
  those with the magnetization that minimizes the free energy~\cite{us_PRL}.
  This procedure is known to minimize the {\em bit error probability} 
  \cite{iba}.  Using the thermodynamic relation $\cF=\cU-{1\ov\be}\cS$, $\be$
  being the inverse temperature (Nishimori's condition corresponds to setting
  $\be\!=\!1$), the free energy of the sub-optimal solutions is given by
  $\cF(m)\!=\!-Fm\m{1\ov\be}\cM(m)$ (for $\cM(m)\!\geq\!0$), while that of the
  correct solution is given by $-Fm_0(p)$ (its entropy being 0).  The selection
  process is explained graphically in
  Fig.\ref{fig:1}(a)-(c). The free energy differences between sub-optimal
  solutions relative to that of the correct solution in the current plots, are
  given by the orthogonal distance between $\cM(m)$ and the line with slope
  $-\be F$ through the point $(m_0(p),0)$.  Solutions with a magnetization $m$
  for which $\cM(m)$ lies above this line, have a lower free energy, while those
  for which $\cM(m)$ lies below, have a higher free energy. Since negative
  entropy values are unphysical in discrete systems, only sub-optimal solutions
  with $\cM(m)\!\geq\!0$ are considered. The lowest $p$ value for which there
  are sub-optimal solutions with a free energy equal to $-Fm_0(p)$ is the
  critical noise level $p_{c}$ for MPM decoding.
  In fact, using the convexity
  of $\cM(m)$ and Nishimori's condition, one can show that the slope $\partial
  \cM(m)/\partial m\!>\!-\be F$ for any value $m\!<\!m_o(p)$ and any $p$, and
  equals $-\be F$ only at $m\!=\!m_o(p)$; therefore, the critical noise level
  for MPM decoding $p\!=\!p_c$ is identical to that of MAP, in agreement with
  results obtained in the information theory community~\cite{MacKay_thrm}.
  
  The statistical physics interpretation of finite temperature decoding
  corresponds to making the specific choice for the Lagrange multiplier
  $\htm\!=\!\be F$ and considering the free energy instead of the entropy.
  In earlier work on MPM decoding in the SP framework~\cite{us_PRL}, negative
  entropy values were treated by adopting different replica symmetry
  assumptions, which effectively result in changing the inverse temperature,
  i.e., the Lagrange multiplier $\htm$. This effectively sets $m\!=\!m_{\e}(p)$,
  i.e. to the highest value with non-negative entropy. The sub-optimal states
  with the lowest free energy are then those with $m\!=\!m_{\e}(p)$.
\end{itemize}
The central point in all decoding schemes, is to select the correct solution
only on the basis of its magnetization. As long as there are no sub-optimal
solutions with the same magnetization, this is in principle possible. As shown
here, all three decoding schemes discussed above, manage to do so. To find
whether at a given $p$ there exists a gap between the magnetization of the 
correct solution and that of the nearest sub-optimal solution, just requires
plotting $\cM(m)(>\!0)$ and $m_0(p)$, thus allowing a graphical determination of
$p_c$. Since MPM decoding is done at Nishimori's temperature, the simplest
replica symmetry assumption is sufficient to describe the thermodynamically
dominant state~\cite{nishimori_book}. At $p_c$ the states with $m_{\e}(p_c)\!=
\!m_0(p_C)$ are thermodynamically dominant, and the $p_c$ values that we obtain
under this assumption are exact.
%
\section{Critical noise level - results}

\begin{figure}[h]
\setlength{\unitlength}{0.6mm}
\begin{picture}(140,60)
\put( 1, 0){\epsfysize=60\unitlength\epsfbox{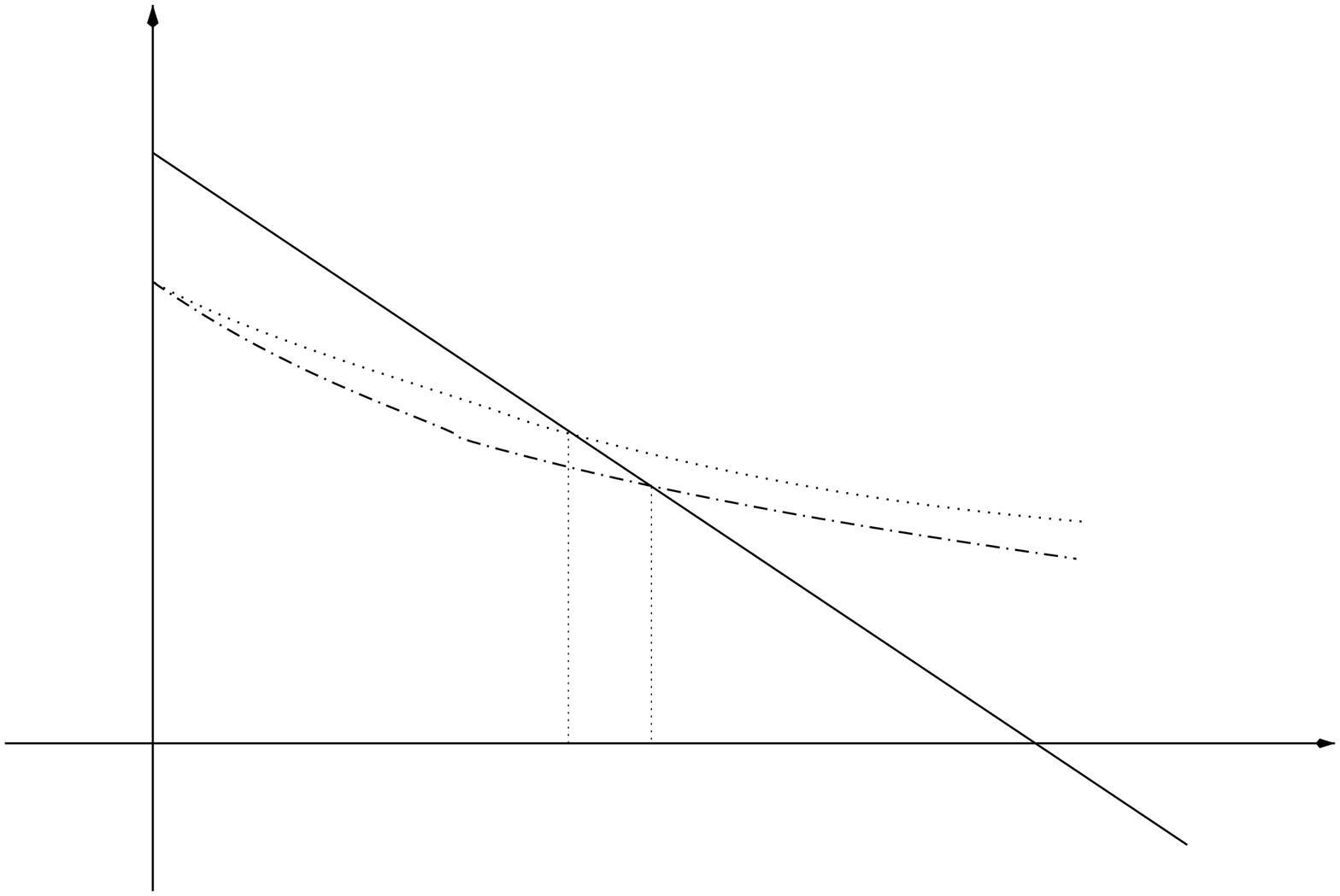}}
\put( 0,55){\mbox{\fns \boldmath ${\rm a)}$}}
\put( 7, 48){\mbox{\fns \boldmath $1$}}
\put(12, 54){\mbox{\fns \boldmath $m$}}
\put(12,  6){\mbox{\fns \boldmath $0$}}
\put(85,  7){\mbox{\fns \boldmath $p$}}
\put(30,  7){\mbox{\fns \boldmath $p_{c,a}$}}
\put(42,  7){\mbox{\fns \boldmath $p_{c,q}$}}
\put(70,  1){\mbox{\fns \boldmath $m_0(p)$}}
\put(72, 27){\mbox{\fns \boldmath $m_{\e,a}(p)$}}
\put(72, 20){\mbox{\fns \boldmath $m_{\e,q}(p)$}}
\put(65,  6){\mbox{\fns \boldmath $0.5$}}
\end{picture}
\put(-43,55){\mbox{\fns \boldmath ${\rm b)}$}}
\put(-35,30){
\begin{tabular}{|c|c|c|c|c|}
\hline
$(K,C)$             & $(6,3)$ & $(5,3)$ & $(6,4)$ & $(4,3)$ \\
\hline
Code rate           & $1/2$   & $2/5 $  & $1/3 $  & $1/4 $  \\
\hline
IT ($\cW_a$)        & 0.0915  & 0.129   & 0.170   & 0.205   \\
\hline
SP                  & 0.0990  & 0.136   & 0.173   & 0.209   \\
\hline
$p_{c,a}$ ($\cM_a)$ & 0.031   & 0.066   & 0.162   & 0.195   \\
\hline
$p_{c,q}$ ($\cM_q)$ & 0.0998  & 0.1365  & 0.1725  & 0.2095  \\
\hline
Shannon             & 0.109   & 0.145   & 0.174   & 0.214   \\
\hline
\end{tabular}
}
\caption{
 {\bf a)} Determining the critical noise levels $p_{c,a/q}$
  based on the function $\cM_{a/q}$, a qualitative picture.
 {\bf b)} Comparison of different critical noise level ($p_c$) estimates.
  Typical set decoding estimates have been obtained via the methods of
  IT~\cite{Aji}, based on having a unique solution to $\cW(m)\!=\!K(m,p_c)$,
 as well as using the methods of SP~\cite{KNM}. The numerical
  precision is up to the last digit for the current method. Shannon's limit
  denotes the highest theoretically achievable critical noise level $p_c$ for
  any code~\cite{Shannon}.
}
\label{fig:2}
\end{figure}

Some general comments can be made about the critical MAP (or typical set) 
values obtained via the annealed and quenched calculations. Since
$\cM_q(m)\leq\cM_a(m)$ (for given values of  $K$, $C$ and $p$), we can derive
the general inequality $p_{c,q}\!\geq\!p_{c,a}$. For all $K$, $C$ values that we
have numerically analyzed, for both annealed and quenched cases, $m_{\e}(p)$ is
a non increasing function of $p$, and $p_c$ is unique. The estimates of the
critical noise levels $p_{c,a/q}$, based on $\cM_{a/q}$, are obtained by
numerically calculating $m_{c,a/q}(p)$, and by determining their intersection
with $m_0(p)$. This is explained  graphically in Fig.\ref{fig:2}(a).
As the results for MPM decoding have already been presented elsewhere
\cite{cactus}, we will now concentrate on the critical results $p_c$ obtained
for typical set and MAP decoding; these are presented in Fig.\ref{fig:2}(b),
showing the values of $p_{c,a/q}$ for various choices of $K$ and $C$ are
compared with those reported in the literature.

From the table it is clear that the annealed approximation gives a much more
pessimistic estimate for $p_c$. This is due to the fact that it overestimates
$\cM$ in the following way. $\cM_a(m)$ describes the combined entropy of $\vn$
and $\vn^o$ as if $\vn^o$ were thermal variables as well. Therefore,
exponentially rare events for $\vn ^o$ (i.e. $m(\vn ^o)\!\neq\!m_0(p)$) still
may carry positive entropy due to the addition of a positive entropy term from 
$\vn$. In a separate study~\cite{KNM} these effects have been taken care of by 
the introduction of an extra exponent; this is not necessary in the current
formalism as the quenched calculation automatically suppresses such
contributions. The similarity between the results reported here and those
obtained in~\cite{reliability} is not surprising as the equations obtained in
quenched calculations are similar to those obtained by averaging the upper-bound
to the reliability exponent using a methods presented originally by Gallager
\cite{Gallager}. Numerical differences between the two sets of results are
probably due to the higher numerical precision here.

\section{Conclusions}
%
To summarize, we have shown that the {\em magnetization enumerator} $\cM(m)$
plays a central role in  determining the achievable critical noise level for
various decoding schemes. The formalism based on the magnetization enumerator
$\cM$ offers a intuitively simple alternative to the weight enumerator formalism
as used in typical pairs decoding \cite{Aji,KNM}, but requires invoking the
replica method given the very low critical values obtained by the annealed
approximation calculation. Although we have concentrated here on the critical
noise level for the BSC, both other channels and other quantities can also be
treated in our formalism. The predictions for the critical noise level are more
optimistic than those reported in the IT literature, and are up to numerical
precision in agreement with those reported in \cite{KNM}. Finally, we have shown
that the critical noise levels for typical pairs, MAP and MPM decoding must
coincide, and we have provided an intuitive explanation to the difference
between MAP and MPM decoding.

\vspace{3mm}
\noindent
{\footnotesize
Support by Grants-in-aid, MEXT (13680400) and JSPS (YK),
The Royal Society and EPSRC-GR/N00562 (DS/JvM) is acknowledged.
}

\end{document}